\input{aipcheck}

\documentclass[
    ,final            
  ]
  {aipproc}

\layoutstyle{6x9}

\begin{document}

\title{Theoretical investigations of isoscalar, isovector and mixed symmetry states in skin nuclei}


\author{N. Tsoneva}{
  address={Institut f\"ur Theoretische Physik, Heinrich-Buff-Ring 16, D-35392 Gie\ss en, Germany},
  email={Nadia.Tsoneva@theo.physik.uni-giessen.de},
  altaddress={Institute for Nuclear Research and Nuclear Energy, 1784 Sofia, Bulgaria}}

\author{H. Lenske}{
  address={Institut f\"ur Theoretische Physik, Heinrich-Buff-Ring 16, D-35392 Gie\ss en, Germany}}


\begin{abstract}
 We present recent investigations on dipole and quadrupole excitations in spherical
skin nuclei, particular exploring their connection to the thickness of the neutron skin.
Our theoretical method relies on density functional theory, which provides us with a proper link between nuclear many-body theory of the nuclear  ground state and its phenomenological description. For the calculation of the nuclear excited states we apply QPM theory. A new quadrupole mode related to Pygmy Quadrupole Resonance in tin isotopes is suggested.  
\end{abstract}

\maketitle


\section{Introduction and theoretical model}
The development of the new experimental facilities on radioactive beams moved forward the nuclear structure physics to the regions of nuclei far from stability \cite{PDRrev:06}.
One of the most interesting results was the discovery of
a new dipole mode at low-excitation energy in nuclei with neutron skin \cite{PDRrev:06}. This bunching of
$1^-$ states resembling spectral structures known to indicate resonance phenomena was named Pygmy Dipole Resonance (PDR). 
The presence of a neutron skin should affect also low-energy quadrupole
states. Due to oscillations of neutrons from the surface region one may expect a formation of a soft quadrupole mode, namely Pygmy Quadrupole Resonance (PQR), which resembles the properties of the PDR. 
  
Here, we present our investigations on the dipole and quadrupole excitations in nuclei from N=50, 82 and Z=50 regions. The approach is based on a Hartree-Fock-Bogoljubov (HFB) description of the ground state \cite{Hof98}
by using a phenomenological energy-density functional (EDF) \cite{Nadia04,Nadia08}. The excited states are calculated with the Quasiparticle-Phonon model (QPM) \cite{Sol76}.
The model Hamiltonian is given by:

\begin{equation}
{H=H_{MF}+H_M^{ph}+H_{SM}^{ph}+H_M^{pp}} \quad
\label{hh}
\end{equation}
which has the structure of the traditional QPM model  \cite{Sol76} but in detail differs in the physical content as discussed in \cite{Nadia08}. The HFB term $H_{MF}=H_{sp}+H_{pair}$ contains two
parts: $H_{sp}$ describes the motion of protons and neutrons in a
static, spherically-symmetric mean-field. $H_{pair}$ accounts for the monopole pairing between isospin identical particles with phenomenologically adjusted coupling constants, fitted to data \cite{Audi95}. The mean-field Hamiltonian compromises microscopic HFB effects \cite{Hof98,Nadia08} and phenomenological aspects like experimental separation energies and, when available, also charge and mass radii. For numerical convenience the mean-field potential is parametrized in terms of (superpositions of) Wood-Saxon (WS)
potentials with adjustable parameters \cite{Nadia08}. 

The remaining three terms present the residual interaction $H_{res}=H_M^{ph}+H_{SM}^{ph}+H_M^{pp}$. As typical for the QPM, we use  separable multipole-multipole $H_M^{ph}$ and spin-multipole
$H_{SM}^{ph}$ interactions both of isoscalar and isovector type in the particle-hole  and multipole pairing $H_M^{pp}$ in the particle-particle channels, respectively. The isoscalar and isovector coupling constants are
obtained by a fit to energies and transition rates of low-lying vibrational states and high-lying GDR's \cite{Nadia08,Vdo83}. 

The nuclear excited states are described by quasiparticle-random-phase- approximation (QRPA) phonons. They are defined as
a linear combination of two-quasiparticle creation and annihilation
operators as follows:
\begin{equation}
Q^{+}_{\lambda \mu i}=\frac{1}{2}{
\sum_{jj'}{ \left(\psi_{jj'}^{\lambda i}A^+_{\lambda\mu}(jj')
-\varphi_{jj'}^{\lambda i}\widetilde{A}_{\lambda\mu}(jj')
\right)}},
\label{eq:StateOp}
\end{equation}
where $j\equiv{(nljm\tau)}$ is a single-particle proton or neutron state;
${A}^+_{\lambda \mu}$ and $\widetilde{A}_{\lambda \mu}$ are
time-forward and time-backward operators, coupling proton and neutron
two-quasiparticle creation or annihilation operators to a total
angular momentum $\lambda$ with projection $\mu$ by means of the
Clebsch-Gordan coefficients $C^{\lambda\mu}_{jmj'm'}=\left\langle
jmj'm'|\lambda\mu\right\rangle$.

The phonons obey the QRPA equation of motion,
which solves the eigenvalue problem, i.e. giving the excitation energies
$E_i$ and the time-forward and time-backward amplitudes
\cite{Sol76} $\psi_{j_1j_2}^{\lambda i}$ and $\varphi_{j_1j_2}^{\lambda i}$,
respectively.

\begin{figure}
\caption{(left) Proton (dash line) and neutron (solid line) ground states densities in Sn isotopes; (right) One-phonon QPM calculations for the total PDR strengths in the
$^{100-132}$Sn isotopes (upper panel) are displayed for comparison
together with the nuclear skin thickness $\delta r$ (lower panel). Experimental data on the
total PDR strengths in $^{116}$Sn and $^{124}$Sn
\cite{Govaert98} and $^{112}$Sn \cite{Ozel} are also shown. In
the lower panel, the skin thickness derived from charge exchange
reactions by Krasznahorkay et al. \protect\cite{Sn-skin1,Sn-skin2}}
\includegraphics[height=.32\textheight]{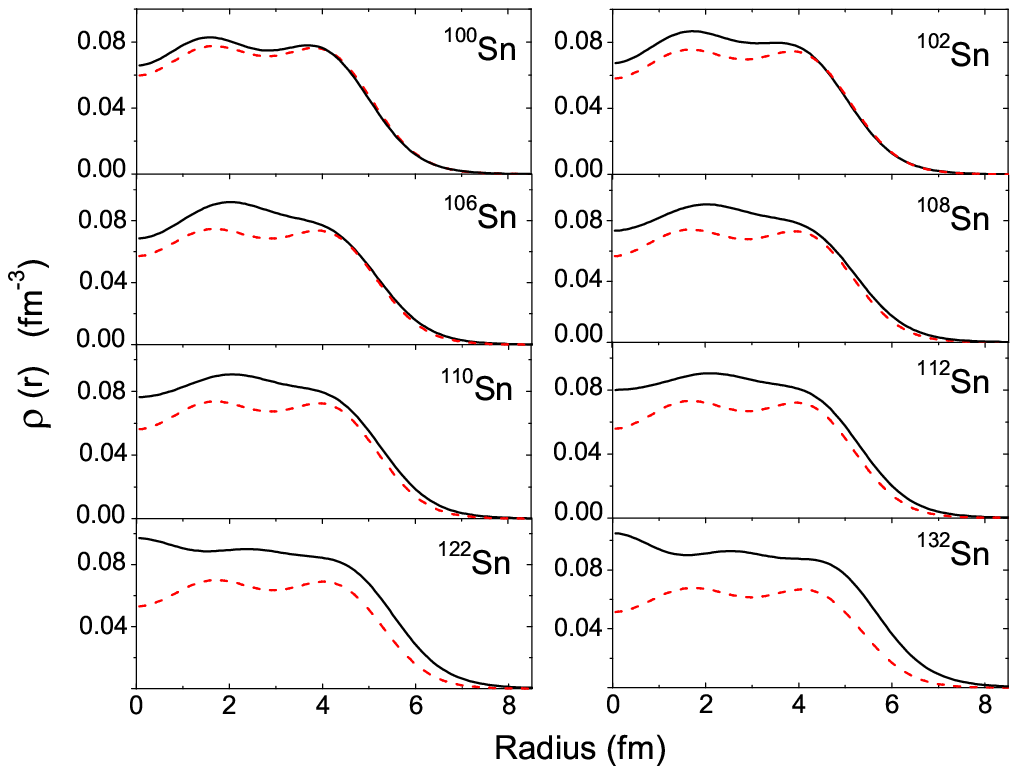}
\includegraphics[height=.32\textheight]{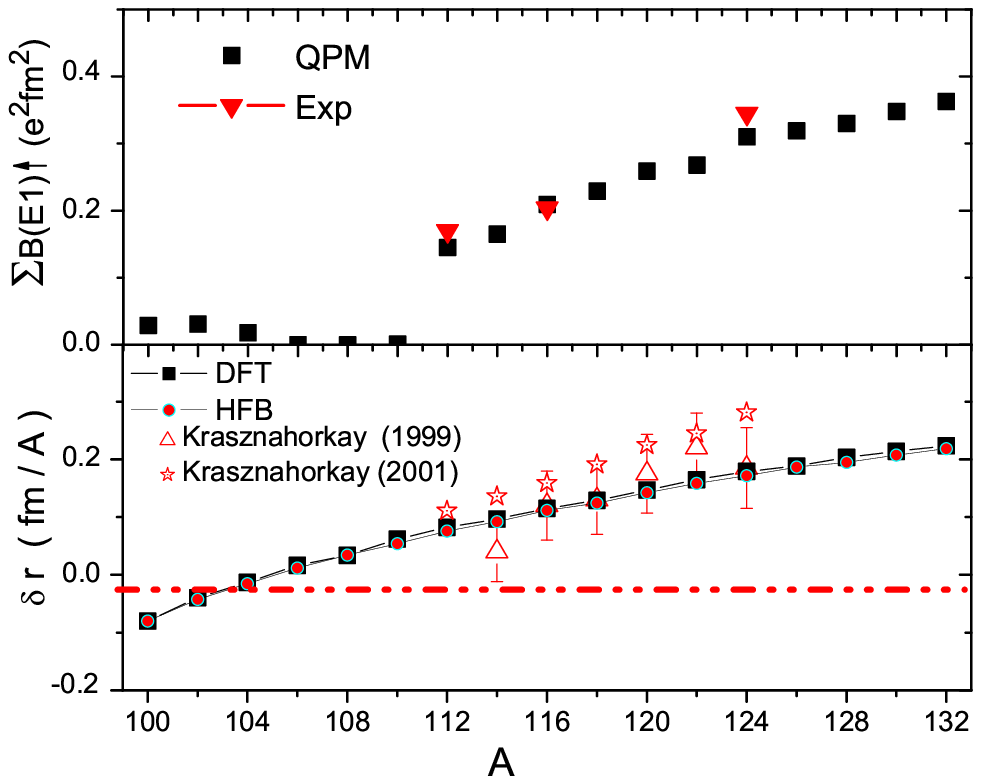}
\end{figure}
\begin{figure}
\caption{(left) QRPA proton (dash line) and neutron (solid line) dipole transition densities summed over the 1$^-$ excited states in a given energy region ( indicated in the figure ) in neutron-rich Sn isotopes and 
(right) in $^{88}Sr$ and $^{90}Zr$, respectively.}
\includegraphics[height=.33\textheight]{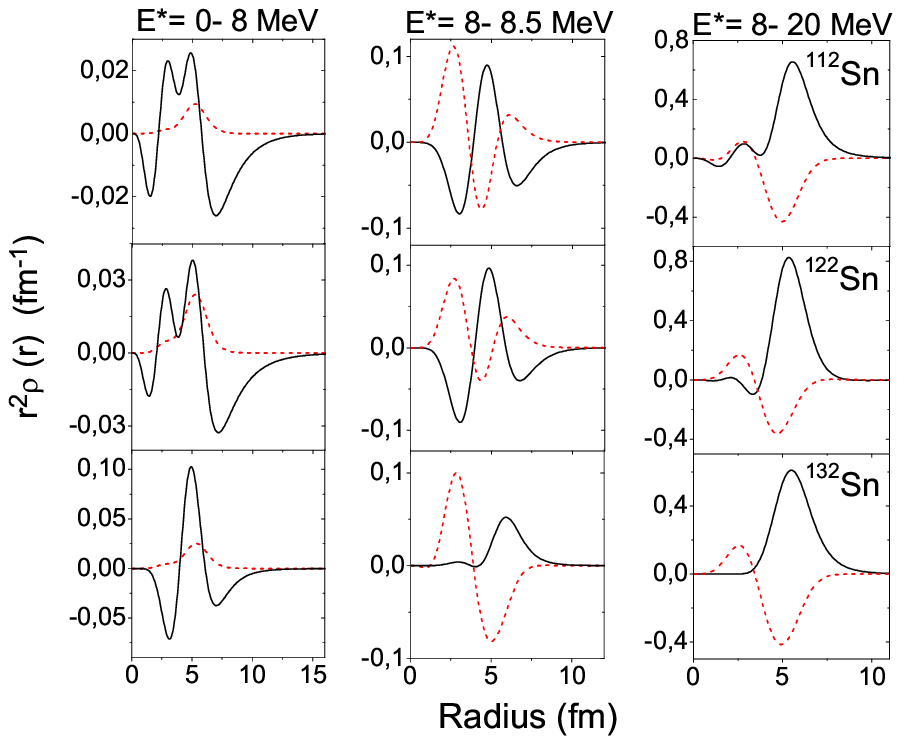}
\includegraphics[height=.3\textheight]{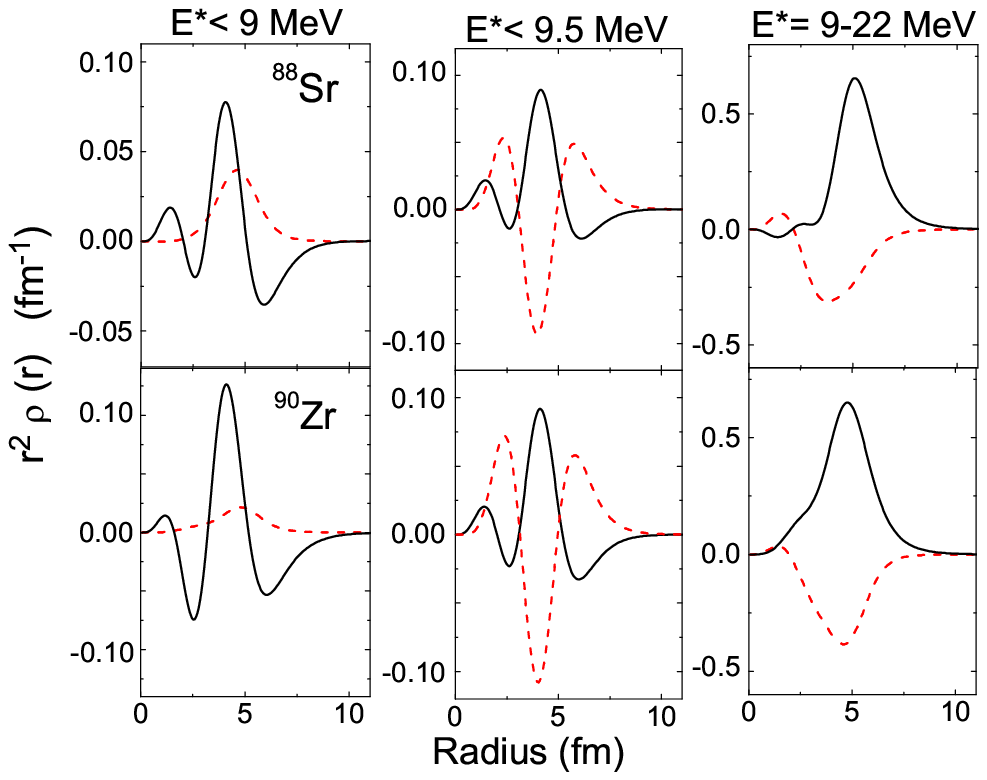}
\end{figure}

As discussed in \cite{Nadia08,Rye:02,Paar} we obtain important additional information from the analysis of the spatial structure of the nuclear response on an external electromagnetic field. That information is accessible by considering the one-body transition densities $\delta\rho (\vec{r})$,
which are the non-diagonal elements of the nuclear one-body density
matrix. The transition densities are obtained by the matrix elements between
the ground state $|\Psi_i\rangle=|J_iM_i\rangle$ and the excited
states $|\Psi_f\rangle=|J_fM_f\rangle$.\footnotetext[1]{ We identify in QPM $|J_iM_i\rangle\equiv |0\rangle$ with phonon
vacuum and obtain the excited states by means of the QRPA state
operator, eq. (\ref{eq:StateOp}), $|J_fM_f\rangle\equiv
Q^+_{\lambda \mu i}|0\rangle$.}
The analytical procedure for the calculation of the transition densities is presented in details in
\cite{Nadia08}.

\begin{figure}
\caption{(left) QRPA calculations of quadrupole states in $^{120}Sn$;
(right) Proton (dash line) and neutron (solid line) quadrupole transition densities summed over the QRPA 2$^+$ excited states in a given energy region ( indicated in the figure ) in $^{120}Sn.$}
\includegraphics[height=.3\textheight]{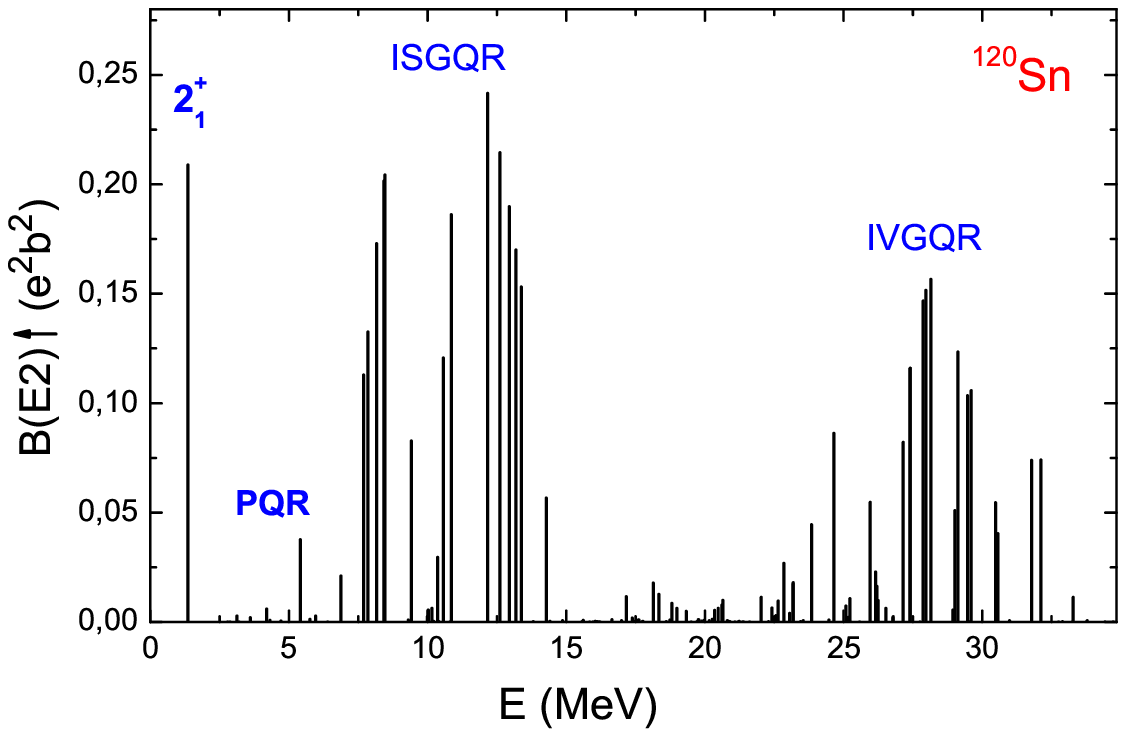}
\includegraphics[height=.3\textheight]{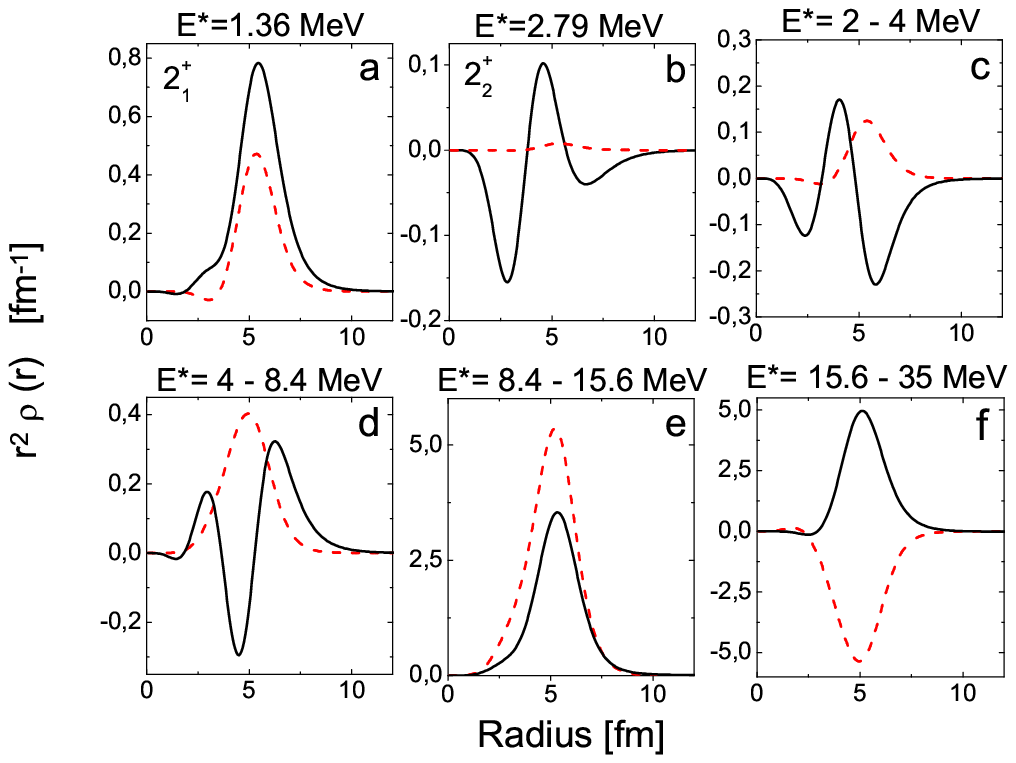}
\end{figure}
\section{Results}
We have pointed out in our previous papers \cite{Nadia04, Nadia08} that the proper determination of the ground state of the nucleus is essential for the extrapolations of the QPM calculations into unknown mass regions. Here, we put special emphasis on a reliable description of the mean-field part, reproducing as close as possible the g.s. properties of nuclei like binding energies, root mean square radii, and separation energies to the required accuracy along an isotopic chain. This is achieved by solving the ground state problem in a semi-microscopic approach \cite{Nadia04, Nadia08}. The calculated ground state neutron and proton densities are displayed in
Fig.1 (left) for several tin isotopes. 
Of special importance for our investigation are the surface regions,
where the formation of a skin takes place. The skin thickness is defined with the deference between neutron and proton rms radii with the equation
$\delta r=\sqrt{<r^2_n>}-\sqrt{<r^2_p>}$ and its evolution with the mass number A for the tin izotopic chain is presented in the lower panel of Fig. 1 (right) (see also in \cite{Nadia08}).
From both figures (Fig. 1 (left) and Fig. 1 (right) lower panel) follow that
for A$\geq$106 the neutron distribution begins
to extend beyond the proton density and the effect continues to
increase with the neutron excess, up to $^{132}Sn$. Thus, these
nuclei have a neutron skin. The situation reverses in $^{100-102}Sn$,
where a tiny proton skin appears. In investigated  N=50, 82 isotones (N$>$Z) the presence of neutron skin was observed as well \cite{Volz06, Ts-Schw08}.

Of special interest are the low-energy dipole excitations in tin  \cite{Nadia04,Nadia08} and N=50, 82 nuclei \cite{Volz06,Ts-Schw08}, which we studied in respect to their connection with skin vibrations. 
From QRPA calculations in $^{112-132}$Sn and N=50, 82 nuclei we obtain a sequence of  low-lying dipole states located below the neutron threshold of almost pure neutron structure. These states exhaust only a tiny fraction, less than 1\%
of the total Thomas-Reiche-Kuhn (TRK) energy weighted dipole sum rule and are related to PDR
\cite{Nadia04,Nadia08,Volz06,Ts-Schw08}.  In general the PDR states are of mixed symmetry and their total transition strength is closely related to the size of the neutron skin  \cite{Nadia08}. The dependence of the calculated total PDR strength on the mass number in $^{100-132}$Sn nuclei is presented in comparison to $\delta$r in Fig.1 (right). The total PDR strength increases when $\delta$r increases and correspondingly the neutron or proton skin thicknesses increase.

The analysis of the dipole transition densities acts as further indicator of PDR and provides a precise separation of the PDR states from the GDR. 
The calculated QRPA proton and neutron dipole transition densities for several Sn isotopes and N=50 isotones (N$>$Z) are presented in Fig.2.
At excitation energies below the neutron emission threshold ( presented in the first column of the left and right plots of Fig.2 ) we observe in-phase oscillation of protons and neutrons in the nuclear interior, while at the surface only neutrons contribute. This picture is associated with PDR mode.
The states above the neutron threshold (second and third column of the left and right plots) indicate isovector oscillations corresponding to the excitation of the GDR. An interesting finding is the most exotic $^{100}$Sn nucleus where the lowest-lying 1$^-$ state has a proton structure. It has been related to a proton PDR and investigated in details by analysis of dipole transition densities \cite{Nadia08}. 

We perform QRPA calculations of 2$^{+}$ states in $^{120}$Sn up to excitation energies E*=35 MeV. Similar to the PDR we observe a sequence of states, located just above the collective 2$^{+}_1$ and below the ISGQR (see Fig.3 (left)) with predominantly neutron structure. The specific admixture of isoscalar and isovector components in their phonon content is a signature of mixed symmetry states (the amplitudes of proton and neutron state vectors are with opposite signs). They are connected with the ground state with small B(E2) transitions.
By analysis of proton and neutron quadrupole transition densities (see Fig 3 (right) bcd ) these states can be related to neutron PQR, where neutrons from the periphery region oscillate against isospin symmetric core.
In addition we can distinguish precisely the PQR states from the collective isoscalar 2$^+_1$ (Fig.3 (right)a) and ISGQR (Fig.3 (right)e). In Fig.3 (right)f QRPA neutron and proton transition densities corresponding to the IVGQR are presented. From the calculation the positions of the maximums for the ISGQR and IVGQR in $^{120}$Sn are determined, which are E*= 12.2 MeV and E*= 28.1 MeV, respectively. These values are close to the given ones in \cite{Hara}.

In conclusion QPM calculations in Z=50 and N=50, 82 nuclei predict low-energy dipole strength in the energy region below the particle emission threshold directly related to the size of the neutron(proton) skin \cite{Nadia08}. The unique character of this mode is associated with PDR,  further confirmed by the shape and structure of the transition densities clearly distinguishable from the GDR. The discussed results are in a good agreement with the experimental data \cite{Nadia08,Volz06,Ts-Schw08}.
In addition quadrupole states were investigated in $^{120}$Sn. By analyzing their structure and proton and neutron transition densities clear separation between collective isoscalar, isovector and mixed symmetry states is achieved. 
A new quadrupole mode related to PQR in $^{120}$Sn, resembling the properties of PDR is suggested. 
Supported by DFG project Le 439/1-4 and BMBF.


\begin{thebibliography}{1}

\bibitem{PDRrev:06}U. Kneissl, N. Pietralla, A. Zilges, J. Phys. G: Nucl. Part. Phys. {\bf 32}, R217-R252 (2006).
\bibitem{Hof98} F. Hofmann and H. Lenske, Phys.~Rev.~{\bf C57}, 2281 (1998).
\bibitem{Nadia04} N. Tsoneva, H. Lenske, Ch. Stoyanov,
Phys.\ Lett.\ {\bf B586}, 213 (2004).
\bibitem{Nadia08} N. Tsoneva, H. Lenske, Phys. Rev. {\bf C77}, 024321 (2008) and refs. therein.
\bibitem{Sol76} V.G.~Soloviev, {\it Theory of complex nuclei}
(Oxford: Pergamon Press, 1976).
\bibitem{Audi95} G.Audi, A.H. Wapstra, Nucl. Phys.~{\bf A595}, 409 (1995).
\bibitem{Govaert98} K.\ Govaert, F.\ Bauwens, J.\ Bryssinck et al.
Phys.\ Rev.\ {\bf C57}, 2229 (1998).
\bibitem{Ozel} B. $\ddot{O}$zel, J. Enders, P. von Neumann-Cosel et al., Nucl. Phys. {\bf A778}, 385-388 (2007).
\bibitem{Sn-skin1}A. Krasznahorkay et al., Phys. Rev. Lett. {\bf 82}, 3216 (1999).
\bibitem{Sn-skin2}A. Krasznahorkay et al., Proc. of the Int. Nucl. Phys. Conf. 2001, by E.  Norman et al.,  editors, American Institute of Physics Proc. No. 610 (New York) 2002, p.751.
\bibitem{Vdo83}A. Vdovin, V.G. Soloviev, PEPAN V.14, N2, 237 (1983).
\bibitem{Rye:02}
N.\ Ryezayeva et al., Phys.\ Rev.\ Lett.\ {\bf 89}, 272502 (2002).
\bibitem{Paar} N. Paar, T. Niksic, D. Vretenar, P. Ring, Phys.~Rev.~{\bf C67}, 034312 (2003).
\bibitem{Volz06}S. Volz, N. Tsoneva, M. Babilon et al., Nucl.~Phys.~{\bf A 779}, 1 (2006).
\bibitem{Ts-Schw08}
R. Schwengner, G. Rusev, N. Tsoneva et al., submitted to PRC.
\bibitem{Hara} M.N.~Harakeh, A. van der Woude, {\it Giant resonances}
(Clarendon press, Oxford, 2001).
  
\end{thebibliography}
\end{document}